# Towards the Internet of Behaviors in airports with a fog-to-cloud approach

Antonio Salis[1][0000-0002-4012-7490]

[1]Engineering Sardegna Srl
Loc. Sa Illetta, SS195 km 2,3 – I-09123 Cagliari, Italy
{antonio.salis}@eng.it

**Abstract.** Recent advances in Internet of Things (IoT) and the rising of the Internet of Behavior (IoB) have made it possible to develop real-time improved traveler assistance tools for mobile phones, assisted by cloud-based machine learning, and using fog computing in between IoT and the Cloud. Within the Horizon2020-funded mF2C project an Android app has been developed exploiting the proximity marketing concept and covers the essential path through the airport onto the flight, from the least busy security queue through to the time to walk to gate, gate changes, and other obstacles that airports tend to entertain travelers with. It gives chance to travelers to discover the facilities of the airport, aided by a recommender system using machine learning, that can make recommendations and offer voucher according with the traveler's preferences or on similarities to other travelers. The system provides obvious benefits to the airport planners, not only people tracking in the shops area, but also aggregated and anonymized view, like heat maps that can highlight bottlenecks in the infrastructure, or suggest situations that require intervention, such as emergencies. With the emerging of the COVID pandemic the tool could be adapted to help in the social distancing to guarantee safety. The use of the fog-to-cloud platform and the fulfilling of all centricity and privacy requirements of the IoB give evidence of the impact of the solution in a smart city environment.

**Keywords.** IoT, IoB, Smart Cities, Cloud Computing, Fog Computing, Fog-to-Cloud orchestration, machine learning, proximity marketing

## 1. Introduction

While the diffusion of the Internet of Things (IoT), as an environment that interconnects an ever-growing number of heterogeneous physical things such as appliances, facilities, vehicles, sensors, etc., to the internet to provide sophisticated applications built with these data [1,2], is continuously proposing new applications and services, the new Internet of Behavior (IoB) has been proposed by Gartner[1] as an extension of IoT, that collects the digital tracks of people lives from a multitude of sources, determining people's attitudes, their interests, preferences and regular habits and practices, and these information could reveal significant information on themselves and can be used to influence their behavior. By 2023, they predict that the individual activities of 40% of the global population will be tracked digitally in order to influence our behavior through feedback loops. That would result in more than 3 billion people, and that by the end of 2025, more than half of the world's population will be subject to at least one IoB programme, whether it be commercial or governmental, to benefit from the knowledge gathered in many commercial, societal, health-related, political scenarios.

---

[1] https://www.gartner.com/smarterwithgartner/gartner-top-strategic-technology-trends-for-2021/

The concept itself is not new as it has been originated in 2012 by Göte Nyman, a famous psychology professor, when he described a way "to offer individuals and/or communities a new means to indicate selected and meaningful behavior patterns, as many as they like, by assigning a specific IB address (analogous to the internet of things) to each behavior pattern just as the person or community sees as best"[2]. Since then, Nyman has clarified his vision, describing the IoB as being the targeting of any ongoing, intended, imagined or planned behavior of people, trying to approach persons at the right moment with appropriate services when such behavior occurs, even if we don't know the identity of such person.

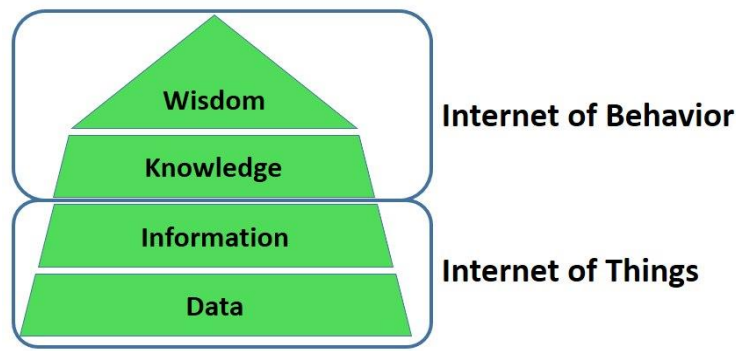

Figure 1 – the DIKW pyramid

While the IoT is concerned with connecting devices, the IoB, leveraging on data analytics and behavioral science, is focused on connecting people and their behaviors, and deals with tools and methods to best use the data to change or influence behaviors. This can be understood looking at the DIKW pyramid in Figure 1: the IoT is more oriented to gather the data from the field and turn it into information, while the IoB is focused on turning that information into knowledge. All this presents some potential ethical concerns depending on objectives and outcomes of the specific uses. The same information that could induce healthy behaviors, thus helping to reduce insurance premiums, could be used to monitor and force purchases. Obviously this would have an impact on the data privacy, and depending on the perception of it, it could reduce the acceptance, adoption and scale of the IoB. So specific features that could provide a trusted environment, with a decentralized processing, with encrypt or anonymize data would be mandatory. To complete the picture, location independence and the ability to operate from anywhere will constitute a major shift in terms of business, requiring a secured distributed cloud processing environment with fast connections, enabling a composable business and leveraging advanced ML/AI technology to enhance the ability to adapt under changing conditions. To support such a challenging shift, the straight "Cloudification" of IoT is problematic, since the approach of transferring all data from the device to the cloud, hosted in remote data centers, generates considerable latency and a large computational load and storage with sensible economic costs. Fog and Edge computing [3], emerged as computing principles where data are processed locally as close as generated, reducing all transmission overhead. By reducing the increase in load on cloud data centers, edge computing can reduce the impact of the increased use of Cloud, better helping people in mobility, while new paradigms as fog computing can help design new technology infrastructures able to process in real-time high volumes of data from the IoT.

---

[2] https://gotepoem.wordpress.com/2012/03/16/internet-of-behaviors-ib/

In the present research, an airport proximity application powered by a managed fog-to-cloud (mF2C) software engine will be described, implementing a IoB solution that preserve the privacy of the end-user, showing that the fog-to-cloud (F2C) approach showcases a full support to the IoB, with better performance than the cloud-only solution.

This manuscript is structured as follows:

- Section 2 introduces the research questions, the Fog-to-cloud approach and the mF2C system developed within the project;
- Section 3 provides a description of the airport use case, its unique proposition, taking advantage of the mF2C platform and fulfilling the IoB concept;
- Section 4 describes in details the deployment in the airport and experimental performance results;
- Section 5 describes the benefits, outcomes and airport managers and ICT/Telco providers' exploitation opportunities;
- Finally, Section 6 describes the relevance of present work, future work and concludes the paper.

## 2. The mF2C system

The EC Horizon 2020 program has funded a new research initiative (mF2C)[3]bringing together relevant industry and academic players in the cloud sector, aimed at designing an open, secure, decentralized, multi-stakeholder management framework for F2C computing, including novel programming models, privacy and security, data storage techniques, service creation, brokerage solutions, SLA policies, and resource orchestration methods [4,5,6,7,8]. The mF2C solution system offers a coordinated management strategy capable of making best use of all existing and potentially available resources in the cloud continuum, from the edge up to the cloud, to execute a service under defined quality constraints. For this the mF2C system proposes a layered and hierarchical architecture, as shown in Figure 2, resources are categorized, using an agent entity to deploy the management functionalities in every mF2C component.

---

[3] http://www.mf2c-project.eu/

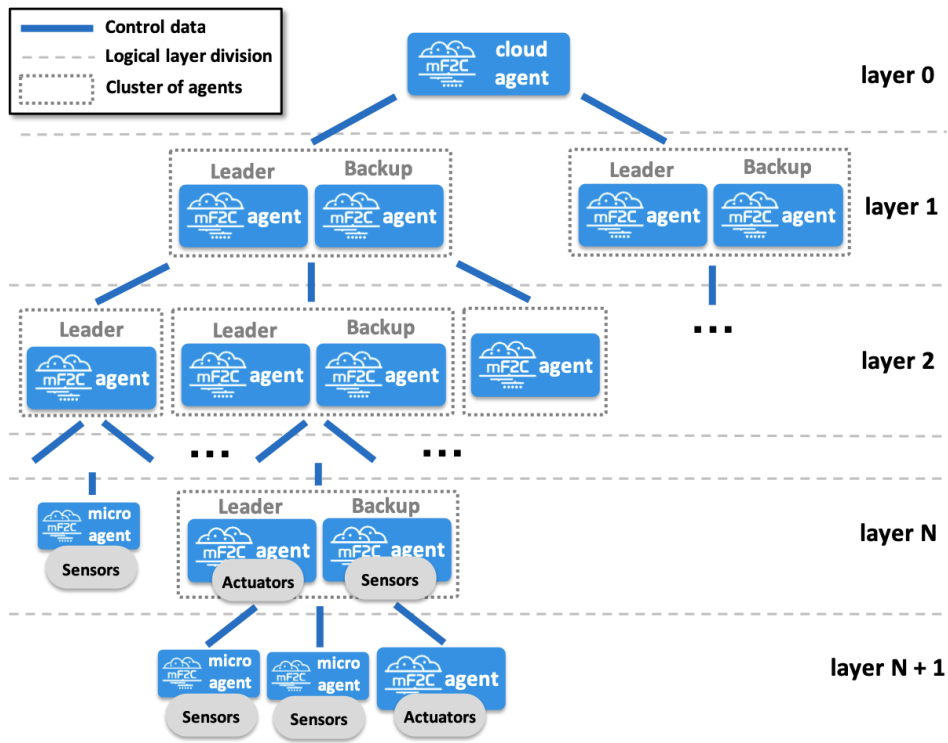

Figure 2 – mF2C hierarchical architecture

The architecture is divided into different logical layers: from layer 0, at the cloud, to layer N+1, at the edge of the network, where three different kind of software entities are deployed: agent, cloud agent and microagent.

The agent is the entity used by default in most of the devices of the architecture, where the cloud agent is an adaptation of the standard agent for the cloud that can be instantiated over one or multiple private or public clouds, and the microagent is a simplified version of the agent designed to be used by edge devices with resource limitations, not able to run a fully operative agent.

Starting from layer 0 (cloud), the instantiation of multiple agents will enable the creation of a layered mF2C architecture, where different agents will be grouped creating multiple clusters, having at least one *leader* (cluster head) and if possible one backup for resilience purposes. In the last layer of every branch, either agents with no devices attached or microagents deployed in highly constrained device could be founds. While microagents can be placed in any layer in the architecture, since it cannot manage other agents, they will act as a leaf in a tree hierarchy.

When an agent receives a request for executing a service, the agent decides the best possible node where it should be executed. If the requested agent has the required resources itself, the service will be executed locally; otherwise it will be forwarded to the leader in the layer above. If the service execution arrives at an agent which controls multiple other agents within lower layers, the agent will act recursively trying to allocate the service using those resources, and if impossible, it will forward the request to the upper layer in the hierarchy.

The cloud agent hosts a directory of defined services, all those that agents can execute. This list of services are reachable by the user logged in through the mF2C dashboard (GUI). The proposed management solution must guarantee that services are executed meeting the required quality of service (QoS) as identified within the Service Level Agreement (SLA) between the user and the provider.

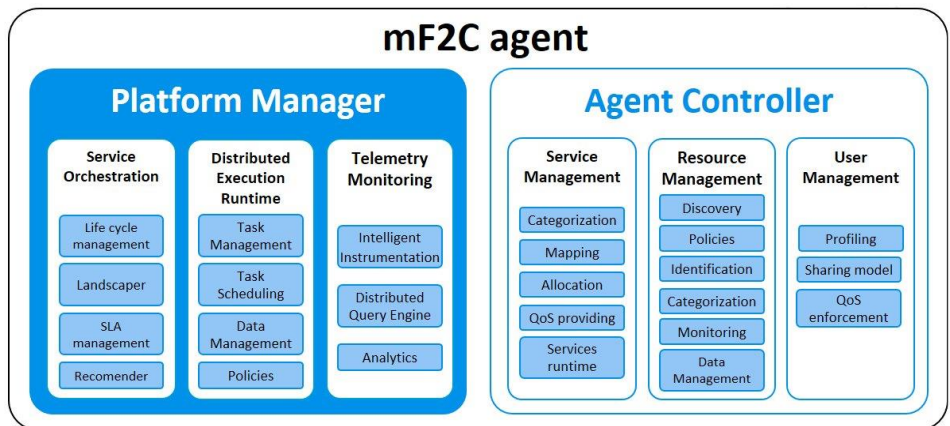

Figure 3: mF2C agent architecture.

Trying to maximize the chances to fulfil the defined SLA, QoS functionalities are split into two different components: i) the QoS providing, enabling the definition of the resources conditions to meet specific QoS requirements and reporting on past SLA violations and; ii) the QoS enforcing that acts at runtime for deploying commands to meet QoS, e.g., reconfiguring resources, services, tasks, etc., on-fly while the service is being executed (an AI-assisted predictor is used for what the delivered QoS will be in runtime). The Resource Manager and Task scheduler are in charge of classifying the available computing nodes and the intelligent task placement according with different objectives, as defined as QoS.

Figure 3 shapes the functional blocks defined for the agent entity, Platform Manager (PM), Agent Controller (AC), Data Management, Security, Event Manager, Graphical User Interface (GUI) and an Application Programming Interface (API) as an entry point. From an implementation point of view an agent is deployed as a collection of Docker[4] containers, with each image exposed via a single REST interface.

The Platform Manager component is an entity acting as a controller for agents in lower layers, and a receiver of control data, when it is being managed by agents from upper layers. It is in charge of service orchestration, telemetry data monitoring from different sources and the coordination of the end-user applications execution. The Agent Controller encloses all functionalities taking care of the resource and user management of local resources, being responsible for defining and executing the assessment of the user's device profile. The role of the Data Management focuses on organizing all mF2C system data resources and offering an interface for accessing this data. The Event Manager is an event tracking component representing a broker that will be used by each of the modules to publish/subscribe to events, e.g., service deployed, device added/removed, etc. Security is provided through three different components, trust (using a Control Area Unit, CAU), web application endpoint security and data protection (using a Security Library with methods for creating message token based on the security level, driven by data classification). The GUI will facilitate users (registration and management) and services operation (registration, catalogue, access, launch).

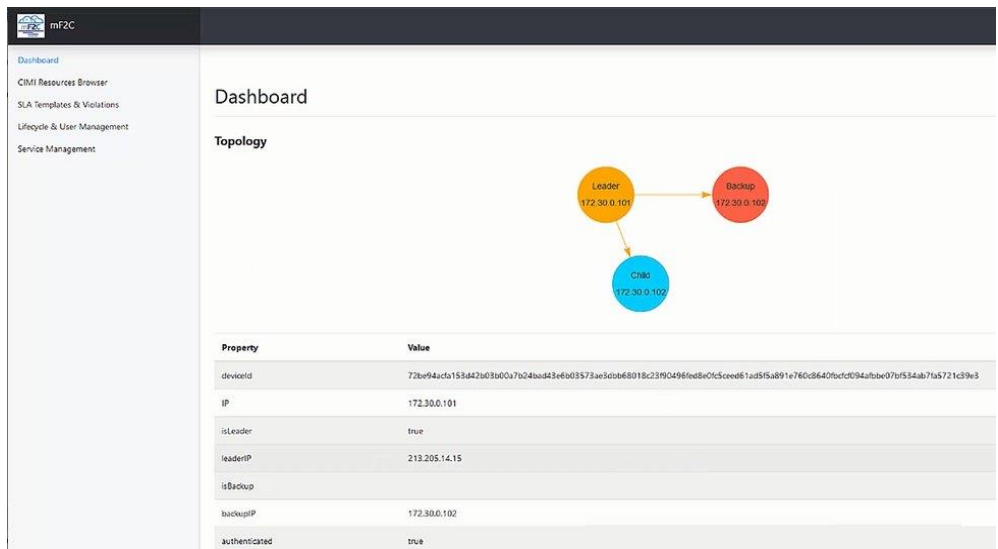

Figure 4 – Dashboard and system topology

Figure 4 shows an example of the mF2C Dashboard browsing the available system topology, with chance to start services, and invoking the mF2C Service Manager. The API of the CIMI[5] module offers the main entry point for all mF2C component.

---

[4] https://www.docker.com/
[5] https://en.wikipedia.org/wiki/Cloud_Infrastructure_Management_Interface

## 3. The Airport Use case

Given the need to spot an environment for IoB exploitation, the analysis has been focused on parts of a smart city, like airports, train stations, hospitals, malls and related parking areas, where there is a concentration of devices, in our case smartphones users, and setup gateways and any other processing elements able to track and engage people in these places, and developing value added services for proximity marketing, with suggestions on best sites to visit, prediction of behavior and movements of consumers, and taking real time decisions, showing in practice the IoB principles.

Looking at the airport field collecting and sharing data on customer behavior can improve the stretch of marketing offering, even with the identity of customers is protected or unknown, using a smart fog gateway embedding cloud connectivity to process large amount of data or request extra data, even data coming from other fogs located in nearby places such as train or main bus station, in order to add value.

The final deployed solution includes a new app based on Android, with an indoor navigator and recommender advisor [9,10,11], driven by machine learning algorithms, providing travelers with a more enjoyable and stress-free experience in the field.

The proposed solution integrates all information that airports already provide through voice announcements, information kiosks and digital monitors, and uses a detailed map of the area, together with the list of available services represented by Points of Interest (POIs), such as restrooms, shops, duty-free areas, information desks, departure gates.

This kind of application is quite different from other offerings [12,13]: most available apps are offered by airlines, but they are limited to their own flights only, and are not able to provide updated information on all departing flights from a specific airport. Google map is based on GPS for people localization, but this does not work well in indoor spaces and does not offer real time information on departing flights. Most smart city applications are based instead in open spaces and use GPS for position tracking, as detaled by Rykowski and Manimuhu [14,15].

The deployed use case has some similarities with the app proposed in the Copenhagen airport[6], but it extends the IoB principles as it can manage merged data and behavior coming from different areas of the smart city.

At the same time the data collected in the fog hub enables an active monitoring of travelers behavior, thus offering benefits to the airport planners as well. Behavioral maps can be showcased both real time and off-line enabling the spotting of bottlenecks in the airport infrastructure, or suggesting ways to better handle emergencies (a passenger being sick, lost children, fire alarm).

### 3.1 Use Case architecture

Figure 5 shows the resulting three layers architecture of the airport system: in the edge layer we have all travelers' Android smartphones using the proposed app, advertised in the airport field by specific totems, and QR code are used for easy download. The edge layer communicate with the access layer represented by eight RaspberryPI3[7], which provide wifi communication and session management, and processing position tracking and proximity application of travelers.

---

[6] https:// www.mapspeople.com/showcases/copenhagen-airport/
[7] https://www.raspberrypi.org/products/raspberry-pi-3-model-b/

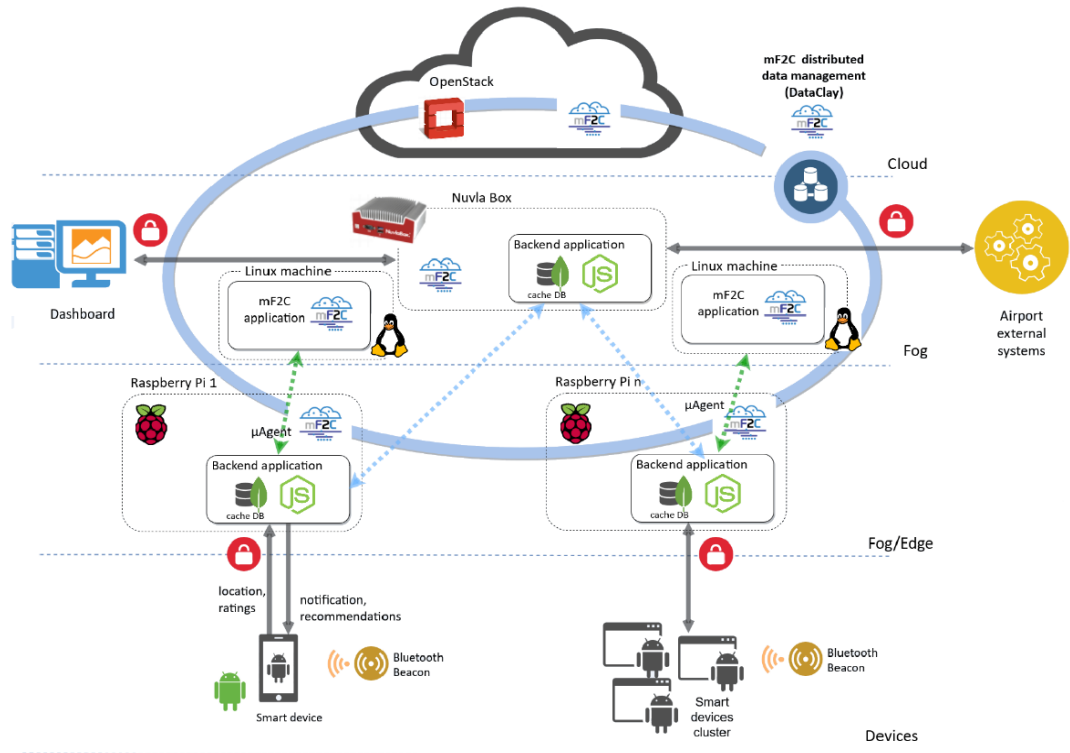

Figure 5: Smart Fog Hub architecture.

The third (fog) layer works as an aggregator, communicating with access nodes and providing real-time computing and storage resources to the edge elements, manage proximity events, using a cached recommender data, and support the admin dashboard with relevant reports. It manages an interface with external airport services, thus collecting airport real time flight events and information. This layer is based on a NuvlaBox[8] appliance playing the role of the fog aggregator and communicates with the fourth (cloud) layer, which is run in a remote datacenter, fiber connected with the airport. The cloud layer, based on an OpenStack[9] instance, provides scalable computing power for big data (including AI models) processing system and manages the long term data storage and analysis.

---

[8] http://www.sixsq.com/products/nuvlabox/
[9] https://www.openstack.org/

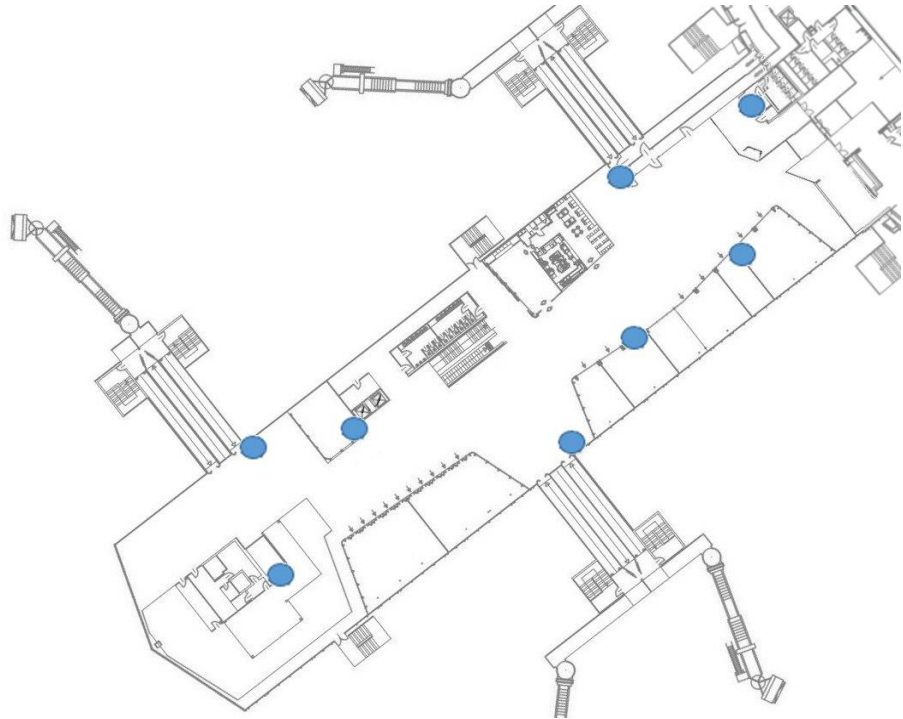

Figure 6: Cagliari airport layout.

All access nodes are positioned in the field to create a regular grid allowing full coverage of wifi signal. Figure 6 represents the topology of the Terminal 1 area, where the installed RaspberryPIs are shown as blue spots. The Android app uses specific trilateral algorithms that evaluate wifi signal strengths to calculate the passenger's position.

The particular positioning of wifi access points and the redundancy supported by the mF2C architecture guarantee optimal use of bandwidth and resilience capabilities, and handover capabilities to link to the strongest signal in the field. The fog aggregator hosts a software component that polls the airport API, so updated flight status data are continuously read from the airport system and distributed to travelers.

The security and privacy of data is guaranteed by the end-to-end mF2C built-in security capabilities [25]: a Certification Authority (CA) running in the cloud node manages a PKI solution. The specific end user is identified assigning it a random Universal Unique Identifier (UUID) and avoiding any hardware code that could give way personal data leaking.

The user keeps the same UUID code unless the app is reinstalled, that would request a new UUID assignment. With this approach the end user could be recognized and managed even while moving across different fog areas, thus enabling the collection of more behaviors and making the ML algorithms more effective.

The adoption of the mF2C system brings two key benefits. First, it enables the scaling up and down of the system, as the number of simultaneous users changes. As the number of users increases, the system manager can deploy more devices in the fog layer with the mF2C agent deploying services on them, thus balancing more efficiently the processing.

At the same time, in case of reduction on the number of users, the system manager could decide to dismiss some resources. As said before an additional benefit from the mF2C usage is its ability to combine more fog areas in the smart city scenario, making the use of IoB more effective.

## 4. Use case deployment

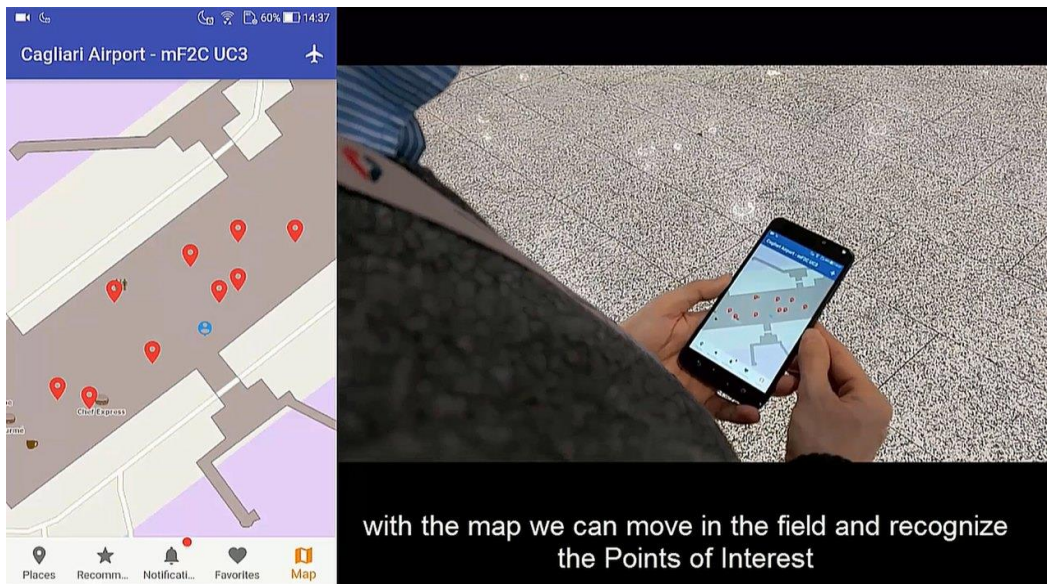

Figure 7: Snapshot of a test session.

The final use case has been deployed in the Cagliari airport. An advertising panel in a totem, located near the entrance of Terminal 1, invited departing travelers to install and use the app. In a time period of four months, until the end of the project, and the beginning of the Covid-19 pandemic, the app has been widely downloaded, installed and used. That allowed us to collect a relevant amount of data for final validation.

Once installed and accepted the privacy terms, the user interface presents the following tabs:
- **Places**: the user can search for interesting POIs from a list of categorized items; selecting a specific POI allows to obtain more details on the offered services;
- **Recommendations**: here the search is driven by ML algorithms that leverage selected topics, user's similarity to other users, and tracked behavior, resulting in a short list of POIs;
- **Notifications**: in this tab main relevant alerts on the flight status, nearby POIs and selected topics are highlighted, where a red spot is used for new notifications;
- **Favorites**: this tab presents most rated POIs from all users, so the user can benefit from the rating of other users while shops can offer special promotions to attract new buyers;
- **Map**: this showcase the airport map with available POIs and real time position of the user while moving around, this is shown in Figure 7;

### 4.1 Performance evaluation

Some tests have been performed in order to validate the system, taking into account performance and responsiveness, with the following measures:
- **Latency**, as measured from the end-user device (smartphone) to the server (fog or cloud device)
- **Response Time**, as the time measured in the client, from the request to the reception of answer.

A laptop has been used to simulate the end-user smartphone, and wi-fi has been used to connect to the access nodes. Jmeter has been used to run batteries of simultaneous client proximity requests to the server, simulating a real world scenario in the airport, thus collecting measures under the different loads.

Different deployments have been done, the proximity calculation has been run using the following server configurations:
1) proximity run on a fog node, so with lower latency (<1msec), but lower processing capacity;
2) proximity run on a remote cloud instance, so higher latency (about 30 msec), but higher processing power;
3) proximity run on one fog node and the remote cloud instance, and the mf2c system has been used for the optimal dispatching of requests;
4) proximity run on two fog nodes and the remote cloud instance, and the mf2c system has been used for the optimal balancing of requests.

The first setup performed well with low number of requests, with the growing number of requests we observed an increase of response time, not fulfilling the real time constraint at the end.

The second setup using the cloud instance presented a quite stable performance that resembles the latency between peer nodes. It is worth noticing that, with the increase of the number of simultaneous requests this setup performed better that the first one.

The third setup uses the mf2c engine with one fog node and the cloud, so it is able to apply the runtime distribution policy, so reaching better performances. With small number of requests most of processing load is run closer to the end-user, while for larger quantities there is an intelligent distribution between the different layers, maintaining an appreciated real time response. We measured an improvement of about 20% compared with the cloud setup.

The fourth setup adds a second fog node, thus enabling a better distribution closer to the end-user, with a further 15% compared with the third setup, with a total improvement of about 35% compared with the cloud setup. The intelligence embedded in the mf2c runtime agent enables the proficient distribution of processing, optimizing the response time, even under severe processing conditions.

Adding more nodes in the fog layer not only facilitate the improvement in performances, but it represent a way to scale the system, while applying the intelligent distribution of processing, moving processing near the end-user, thus saving latency time, and off-load part of processing to the cloud to avoid fog nodes overloading.

The load balancing and intelligent distribution of processing adopted in the mf2c engine differs from similar approaches such as Li [16] and Maia [17]. Li in particular uses a different classification of resources and uses a scheduling approach based on genetic algorithms, which seems less performing than the AI/Deep Learning approach in mF2C, that performs better in more dynamic scenarios even with devices at the edge.

Petri and Sinaeefourfard [18,19] describe in more detail the different strategies for off-loading in Centralized-to-Decentralized and Centralized-to-Decentralized, and which are the scenarios where each example fits best.

## 5. Benefits and Outcomes

The airport application with the support of the Smart fog hub system has been designed from the very beginning, with the goal to demonstrate the IoB capabilities of tracking and engaging interested people in the airport area and use a machine learning based advisor to provide suggestions on the best way to use available services, achieving an outstanding customer experience. We succeeded in demonstrating that the fog-to-cloud criteria enables a more efficient implementation of real-time advisory services in proximity. In particular the following business improvements have been reached:

- **IoB services based on proximity in a smart city scenario:** the increasing number of travellers that install and use the Android app demands processing distribution capabilities starting from the edge nodes where data has been generated, thus optimizing the requirements of fast response demanded by the application. The mF2C orchestration module played a key role in supporting SLA policy definition and enforcement at runtime. This feature enables the delivery of personalized offers to the customers, according to their preferences and behavior;
- **Use of ML to advise traveller:** Machine Learning capabilities have been embedded in the application enabling the foreseen IoB capabilities, at the same time similarities between users have been used to propose more recommendations, with consequent benefits;
- **Embedding of all information on flights in the app:** the application collects real-time information available on flight in the terminal area, making them available according to the traveler's expressed preferences and needs;
- **Security and Privacy:** the application makes full advantage of the security and privacy by-design enabled capabilities provided by the mF2C system, to guarantee full GDPR compliancy, and in case anonymizing information as long as the IoB features perform as expected, and full acceptance of the IoB oriented features by the users is ensured;
- **Fog computation:** The extensive use of the new fog-to-cloud paradigm, that pushes the processing closer to where data is produced and needed, offers better performances and control on managed data. The Fog Hub in the airport plays a major role in this, improving local processing and data storage, and using the cloud only for huge long term big data processing;
- **Administrative portal for overall control and management**: the deployment of the administrative portal provides a better tool to airport planners and managers to monitor the overall situation in the terminal area. The dashboard offers both static and dynamic reports, such as graphical diagrams on users' behavior that showcase the use of available resources, waiting times in different gates or security checks, thus facilitating the spotting of bottlenecks in the infrastructure;

- **Use of Serverless to improve the efficiency at the edge**: the redesign of business processes of main services as microservices, as shown in Figure 8, has enabled the deployment of smaller chunks of code with Docker, and the Serverless capabilities supported by the mF2C agent made possible to run more software components at the edge with better overall performance.

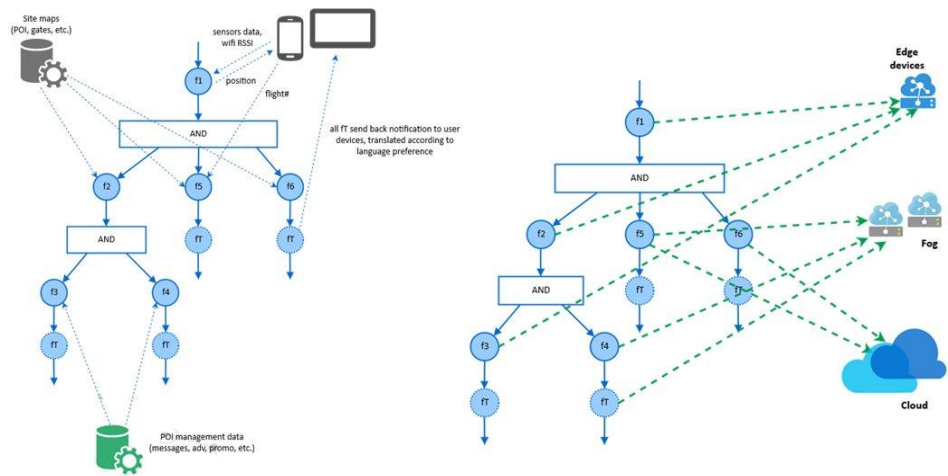

Figure 8: design oriented to set of microservices

For sure the deployment of the mF2C system and all the airport features and capabilities listed above demonstrated interesting business opportunities, but more relevant, the hierarchical structure of the mF2C makes easy to merge different fog areas through the cloud, and let them works together, with the mF2C acting as the glue that interconnects all system components [24].

So a complex scenario like the smart city can be split in several fogs (airport, train station, harbor, shopping centers, hospital, etc.) with a "divide-et-impera" approach [20,21], leveraging on the pillars of interoperability, mobility, fast response, adaptive and autonomous processing, as shown in Figure 9. This could leverage the identity management capabilities to merge all behaviors of users making possible to produce customized recommendations and proposals, thus improving both the customer experience and the effectiveness of marketing proposals.

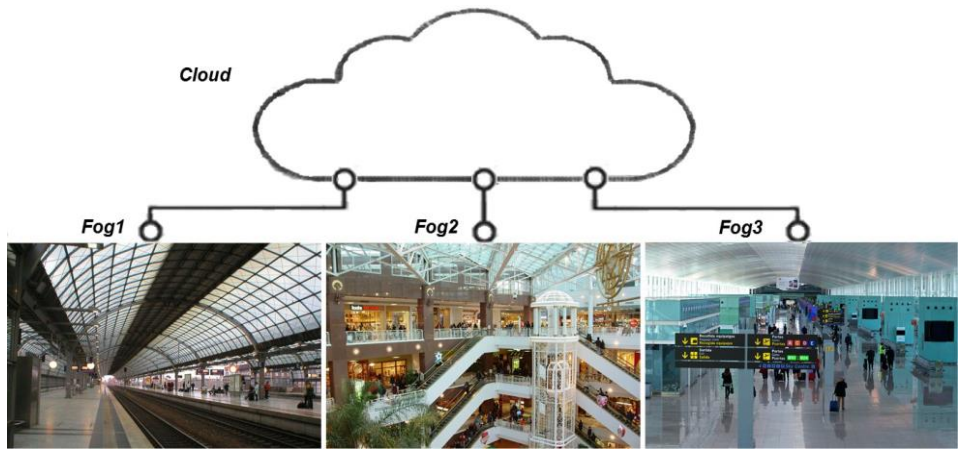

Figure 9: Managing multiple fogs in a smart city scenario

In terms of potential exploitation of the project outcomes, as the airport traffic is continuously growing, airport managers are worried about checking that the infrastructure successfully support this traffic. In this scenario the demonstrated tools fulfils a practical need, dynamically monitoring the area and making possible the extension of the infrastructure using a data-driven approach.

The recent shock caused by the COVID pandemic impacted also the airport areas, so the social distancing enforcement has emerged as an additional requirement to be enforced, so dynamic detection of people clusters (as in Figure 10) and avoiding people clustering [22], sticking in the limits imposed by law, has come to be very important.

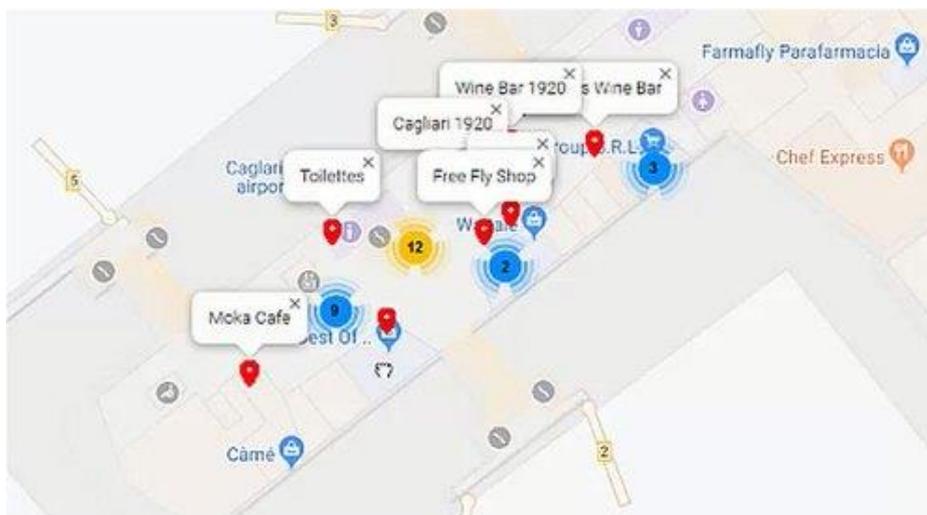

Figure 10: Snapshot of people tracking and clustering

It has been easily determined that the position tracking could drive suggestions in this perspective, so if a shop has reached the maximum allowed number of customers, the traveller could be advised about less busy alternatives, and some virtual queues could be setup to alert interested people when space is available and it is their turn. The same logic could be applied to manage emergency cases such as fire alarms: very short advices could be provided through the app preventing panic behaviors.

## 6. Conclusions

The Smart Fog Hub in airport showed to be a very novel and effective solution that demonstrates all the IoB benefits and potential impact, making use of the fog-to-cloud approach to deliver efficient services in a smart city scenario. The data-driven approach derived from the IoT is the perfect enabler for the IoB adoption, at the same time it is also possible to merge user's data coming from different fog areas in the smart city, thus boosting the IoB effectiveness. It is worth to remind even the great capabilities to make the smart city safer even in case of pandemics, while inducing some safer behaviors to citizens. The other side of the coin is related with the privacy and security aspects that the IoB impact: every solution should be built by-design with all privacy and security aspects managed properly. Proper rules for privacy and security should be organized from the edge, where the data owners are and where data will be generated: this is key for GDPR compliance and further user acceptance, as in the described application to the use case in the airport. In all cases an anonymization to sensitive data still offers opportunities for a successful use of the IoB in smart cities. We plan to follow up this work in future research projects, and the emerging IDSA framework [23] will be investigated as it aims to define a global standard that secures the exchange of data in compliance with major privacy and security requirements.

**Acknowledgments** This work is supported by the H2020 mF2C project (730929). The author wish to thanks prof. Xavi Masip-Bruin, Eva Marin-Tordera (Università Politècnica de Catalunya), Rosa M. Badia (Barcelona Supercomputing Center) and Jens Jensen (Science Technology Facility Council/UK Research and Innovation) for their valuable suggestions.

This work is dedicated to the memory of my beloved parents Franco and Chicca and dear uncle Cicci, as sources of inspiration.